\documentclass[twocolumn,twoside,slac_two]{revtex4}
\usepackage{graphicx}
\usepackage{fancyhdr}
\usepackage{hyperref}
\hypersetup{
    colorlinks=true,
    urlcolor=magenta,
}
\begin{document}

\title{Radio detection of Extensive Air Showers}

%

\author{Frank G. Schr\"oder}
\affiliation{Institut f\"ur Kernphysik, Karlsruhe Institute of Technology (KIT), Germany}

\begin{abstract}
Detection of the mostly geomagnetically generated radio emission of cosmic-ray air showers provides an alternative to air-Cherenkov and air-fluorescence detection, since it is not limited to clear nights. 
Like these established methods, the radio signal is sensitive to the calorimetric energy and the position of the maximum of the electromagnetic shower component. 
This makes antenna arrays an ideal extension for particle-detector arrays above a threshold energy of about 100 PeV of the primary cosmic-ray particles. 
In the last few years the digital radio technique for cosmic-ray air showers again made significant progress, and there now is a consistent picture of the emission mechanisms confirmed by several measurements. 
Recent results by the antenna arrays AERA and Tunka-Rex confirm that the absolute accuracy for the shower energy is as good as the other detection techniques. 
Moreover, the sensitivity to the shower maximum of the radio signal has been confirmed in direct comparison to air-Cherenkov measurements by Tunka-Rex.
The dense antenna array LOFAR can already compete with the established techniques in accuracy for cosmic-ray mass-composition. 
In the future, a new generation of radio experiments might drive the field: either by providing extremely large exposure for inclined cosmic-ray or neutrino showers or, like the SKA core in Australia with its several 10,000 antennas, by providing extremely detailed measurements. 
\end{abstract}

\maketitle

\thispagestyle{fancy}

\section{Introduction}
Antenna arrays provide an alternative way for the measurement of high-energy cosmic-ray air showers initiated by any type of primary particle \cite{SchroederReview2016, HuegeReview2016}.
The threshold of current radio arrays is around $10^{17}\,$eV.
At these energies, however, no neutrinos and gamma rays have been discovered, yet. 
Thus, apart from the search for these neutral particles, the main instrumentation goal is to increase the measurement accuracy for the composition of the primary cosmic nuclei as a function of energy. 
This increase in accuracy is crucial to distinguish competing scenarios for the transition from Galactic to extragalactic cosmic rays expected in this energy range \cite{KGheavyKnee2011, 2013ApelKG_LightAnkle}, and to understand the origin of the most energetic Galactic and extragalactic particles \cite{AugerUpgradeICRC2015}. 
Moreover, a higher precision is required to study the weak anisotropy of the primary cosmic rays separately for different mass groups \cite{AnisotropyComposition2011, LargeScaleAnisotropySearch2015}. 
Radio extensions of particle-detector arrays can bring this desired enhancement of precision and accuracy.

\begin{figure*}
  \centering
  \includegraphics{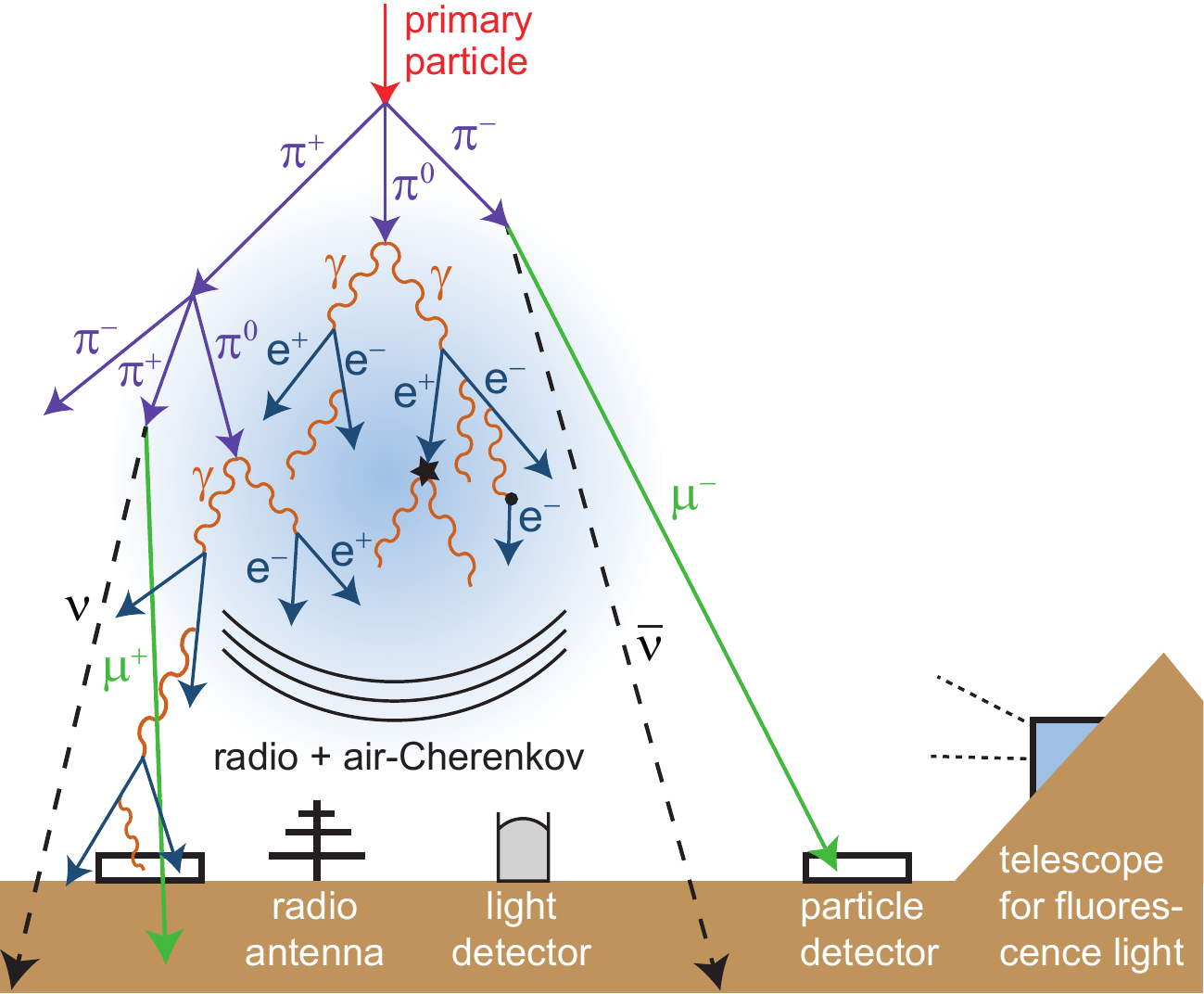}
  \caption{Simplified scheme of a cosmic-ray air shower initiated by a nucleus. 
  While muons are measures primarily with particle detectors, the electromagnetic shower component is also measured indirectly by many experiments via emitted Cherenkov and fluorescence light during night, and at any time via its radio emission \cite{SchroederReview2016}.}
  \label{fig_showerSketch}
\end{figure*}

A traditional method for air-shower detection is particle detectors on and under ground (see figure~\ref{fig_showerSketch}). 
They can directly detect muons and particles of the electromagnetic cascade (electrons, positrons, photons) of the air shower. 
The ratio between the number of electrons and the number of muons provides an estimator for the composition of the primary particles, since showers initiated by heavy nuclei on average contain more muons than showers initiated by light nuclei, and photon induced showers have almost no muons. 
Unfortunately, the interpretation of such particle measurements suffers from a deficient description of the atmospheric particle cascades.
Since the first interactions in the cascades occur at energies far beyond the range probed at accelerators, one extrapolates from experimentally proven knowledge, and even the best hadronic interaction models fail to describe the number and distributions of muons correctly \cite{KASCADEGrandeMuons2015, AugerMuonsInclinedPRD2015}.

The electromagnetic component of air showers can be described more accurately, since the underlying physics is better understood than for the hadronic component producing the muons. 
Traditionally optical methods are used for measurement of the electromagnetic component: air-Cherenkov light emitted in the forward direction of the shower and fluorescence light emitted isotropically by the air after excitation of nitrogen molecules by the traversing shower.
Unlike the muons, measurements of the electromagnetic component are in agreement with simulations at all accessible energies until about $10^{20}\,$eV.
This reduces systematic uncertainties in the interpretation of optical measurements significantly compared to particle measurements \cite{KampertUnger2012}. 
The main disadvantage of the optical methods is their low duty cycle, since optical measurements require dark and clear nights. 

Antenna arrays combine the two main advantages of particle and optical detection: radio detection of air showers is possible under almost any weather and light conditions except for thunderclouds directly over the array, since these alter the radio signal \cite{2011ApelLOPES_Thunderstorm, SchellartLOFARthunderstorm2015}. 
Moreover, the radio signal provides a measure of the well understood electromagnetic component. 
As an additional advantage, to current knowledge the emission and propagation of the radio signal depends less on atmospheric conditions than Cherenkov or fluorescence light. 
This can make radio detection even more accurate than the optical techniques, and current radio arrays have achieved similar accuracies of $15-20\,\%$ for the energy \cite{2014ApelLOPES_MassComposition, TunkaRex_Xmax2016, AERAenergyPRL2015}.
The accuracy for the atmospheric depth of the shower maximum, $X_\mathrm{max}$, can be as good as $20\,$g/cm$^2$ depending on the experiment \cite{BuitinkLOFAR_Xmax2014, 2014ApelLOPES_MassComposition, TunkaRex_Xmax2016}.
Together with the electron-muon ratio $X_\mathrm{max}$ is the most important estimator for the mass composition of the primary cosmic rays. 

For significant further progress in cosmic-ray physics, each single mass estimator alone probably is insufficient, and the combined accuracy of both mass estimators, i.e., $X_\mathrm{max}$ and the electron-muon ratio, might be required. 
Since the radio signal is sensitive to $X_\mathrm{max}$ and to the size of the electromagnetic component, this makes radio detection an ideal complement to muon detectors \cite{Holt_TAUP2015, SchroederAERA_PISA2015}. 
Furthermore, this combination automatically compensates for an apparent disadvantage of the radio technique, since the muon detector can trigger the readout of the radio antennas. 
Artificially generated radio disturbances can have a signature similar to the pulses emitted by air showers and are difficult to separate when using simple radio detectors. 
Although self-triggered radio detection of air showers has been proven possible \cite{RAugerSelfTrigger2012, TREND2011, ARIANNA_ARENA2016}, achieving high purity requires radio quiet sites.
Thus, it is much simpler to use a co-located particle detector as a trigger, as done by many current radio arrays. 

In summary, the radio technique is about to cross the threshold from proof-of-principle demonstrations to a serious contribution to cosmic-ray physics. 
Large-scale extensions of particle-detector arrays by radio antennas will be economic and most useful for all scientific goals requiring more accurate measurements.

\begin{figure*}
  \centering
  \includegraphics[width=0.533\linewidth]{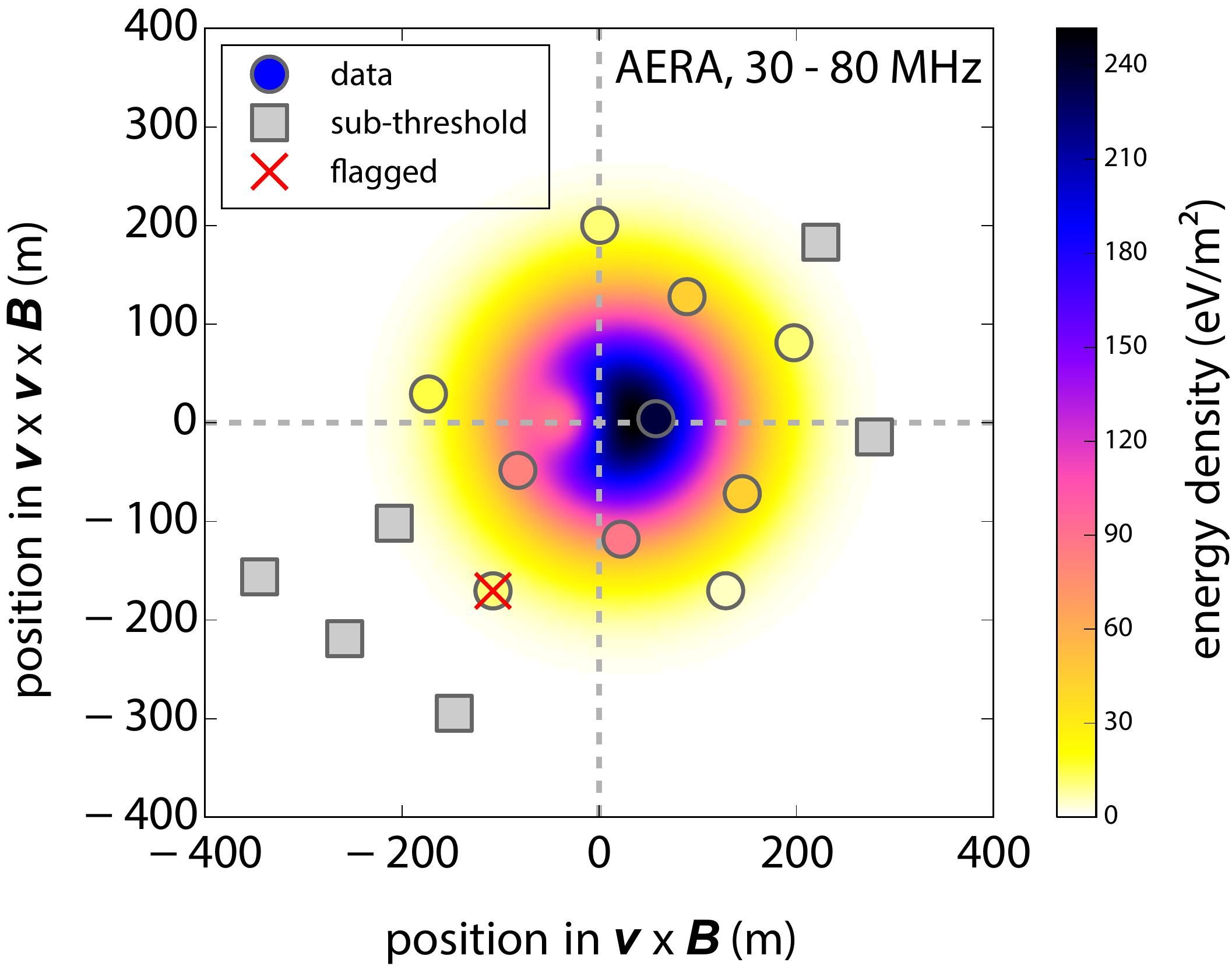}
  \hfill
  \includegraphics[width=0.45\linewidth]{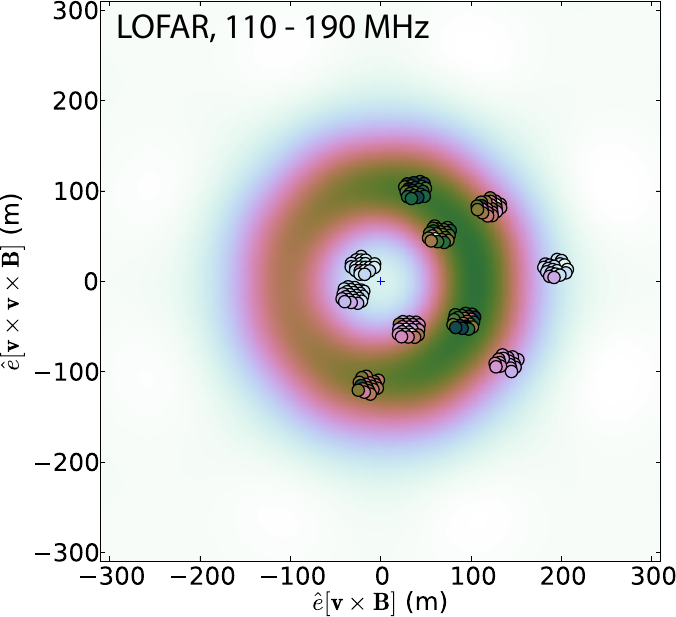}
  \caption{Footprint of the radio signal in the shower plane in two different frequency bands: left an AERA event measured at $30-80\,$MHz \cite{GlaserICRC2015}, right a LOFAR event measured at $110-190\,$MHz \cite{NellesLOFAR_CherenkovRing2014}. 
  The color in the circles is the measured signal strength, the surrounding color is a fit of the radio footprint to these measurements. 
  The footprint is slightly asymmetric due to the interference of the geomagnetic and Askaryan effects. 
  The coordinates are along the Lorentz force ($v \times B$) and orthogonal to it ($v \times v \times B$), where $v$ is the direction of the shower axis and $B$ the geomagnetic field.}
  \label{fig_footprints}
\end{figure*}

\section{Properties of the radio emission}

The radio signal of air showers is coherent, forward-beamed emission generated predominantly by the electrons and positrons in the electromagnetic cascades. 
This emission arrives as a short pulse with a width between about one and a few tens of nanoseconds depending on the distance to the shower axis. 

On first impression the radio signal and the Cherenkov-light emitted by air showers have similar properties: 
both are beamed in the forward direction with an opening angle of the beam of the order of $3^\circ$, i.e., the footprint on ground typically has a diameter of the order of $200\,$m for vertical showers. 
However, there are several differences between the radio signal and the Cherenkov light emitted by air showers.
First, the radio emission is coherent. 
The field strength (= amplitude) of the radio signal increases approximately linearly with the number of electrons in the electromagnetic cascade, which is roughly proportional to the shower energy. 
The radiation energy and intensity consequently scale quadratically with the shower energy, contrary to Cherenkov light whose intensity increases only linearly with the shower energy. 
Second, the shape of the radio footprint on the ground depends on the frequency band. 
The Cherenkov-like ring on ground becomes sharper for higher radio frequencies, since smaller wavelengths require stricter coherence conditions for the emission.  
While the ring is completely filled with signal at frequencies below $100\,$MHz, at higher frequencies the emission is detectable only at the ring (see figure \ref{fig_footprints}) \cite{NellesLOFAR_CherenkovRing2014}, which is visible up to at least a few GHz \cite{CROME_PRL2014}. 

This implies that antenna arrays operating below $100\,$MHz can be relatively sparse with spacings up to $200\,$m for vertical showers and spacings of larger than $1\,$km for inclined showers \cite{AERAinclined_ARENA2016}. 
At frequencies of a few GHz the spacing would have to be of the order of $10\,$m, since the width rather than the diameter of the Cherenkov-like ring will be relevant for efficient detection.
The advantage at GHz frequencies is a much lower external background. 
While at MHz frequencies the hardly reducible Galactic background covers the radio signal of air showers at energies below $10^{17}\,$eV, at GHz frequencies only internal background of the receiver counts.
Thus, the threshold could be one or two orders of magnitude lower for dense GHz arrays. 
However, the requirement of a much denser antenna spacing makes GHz arrays too expensive at the present, which is the principle reason why current experiments mostly operate at a typical frequency band of $30-80\,$MHz. 

\begin{figure*}
  \centering
  \includegraphics[width=0.6\linewidth]{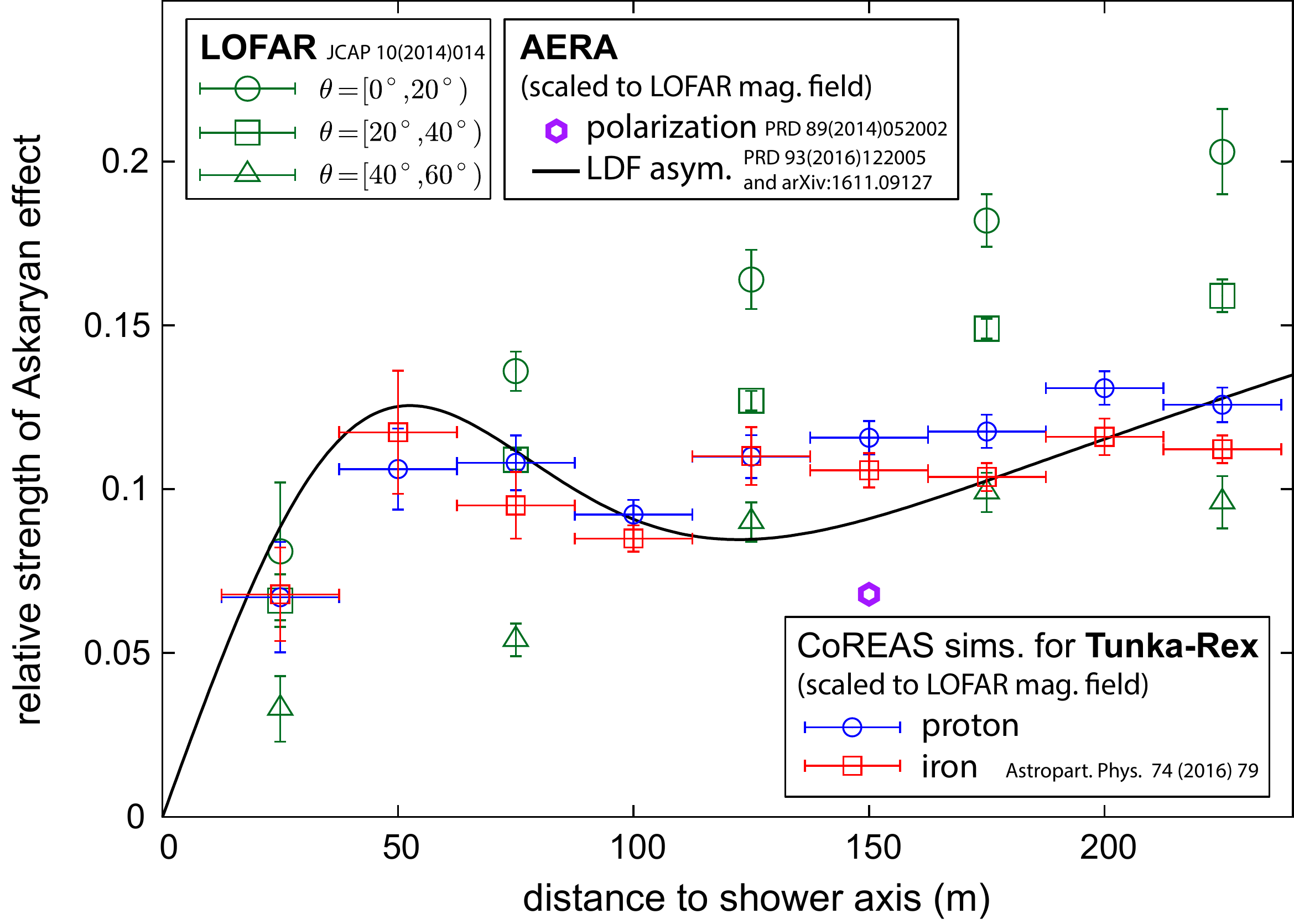}
  \caption{Strength of the Askaryan emission in relation to the geomagnetic one for a geomagnetic angle $\alpha=90^\circ$; LOFAR: polarization measurements \cite{SchellartLOFARpolarization2014}; AERA: polarization measurements of prototype setup \cite{AugerAERApolarization2014} + asymmetry of the radio footprint implicit in the lateral distribution function (LDF) \cite{AugerAERAenergy2015, Kostunin_ARENA2016}; Tunka-Rex: CoREAS simulations \cite{KostuninTheory2015}; from \cite{SchroederReview2016}.}
  \label{fig_asymmetry}
\end{figure*}

Third, another important difference between radio and Cherenkov light are the physics processes causing the emission. 
While the Cherenkov light is caused by the speed of the electrons and positrons being faster than the speed of the light in the atmosphere, the radio emission originates from two completely different mechanisms. 
The dominant mechanism is the geomagnetic deflection of electrons and positrons in opposite directions by the geomagnetic field.
This leads to linearly polarized radio emission pointing in the direction of the geomagnetic Lorentz force, $\bf{v} \times \bf{B}$, with $\bf{v}$ the shower axis and $\bf{B}$ the geomagnetic field.
The strength of the geomagnetic emission is approximately proportional to the geomagnetic Lorentz force and, thus, to the sine of the 'geomagnetic angle' between the shower axis and magnetic field of the Earth, $\sin \alpha$.
A generally weaker, but not negligible contribution to the radio signal originates from the Askaryan effect.
That is emission due to the time-variance of the negative net charge of the shower front (the total charge of the Earth is conserved, since the atmosphere is ionized and charged positively by the traversing shower). 
The Askaryan emission is radially polarized around the shower axis.
Depending on the azimuthal location around the shower axis it either adds constructively or destructively to the geomagnetic emission. 

Other emission mechanisms have been proposed, but have not yet been experimentally confirmed - except of the acceleration of electrons by atmospheric electric fields which is important only during thunderstorms \cite{2011ApelLOPES_Thunderstorm, SchellartLOFARthunderstorm2015}. 
Moreover, simulation codes relying on geomagnetic and Askaryan emission consistently describe various measurements within an experimental scale uncertainty of about $20\,\%$ \cite{2015ApelLOPES_improvedCalibration, TunkaRex_NIM_2015}.
They also describe the dependencies on the distance to the shower axis and on the arrival direction sufficiently well \cite{2013ApelLOPESlateralComparison}. 
This means that any potential mechanism not yet discovered likely contributes to the total emission by much less than $20\,\%$.
For further reading on the theoretical understanding of the radio emission by air showers please refer to references \cite{Werner_EVA2012, deVries2013, HuegeTheoryOverview_ARENA2012}, and for the recent simulation codes CoREAS and ZHAires to references \cite{HuegeCoREAS_ARENA2012, Alvarez_ZHAires_2012}.

Due to the interplay of the geomagnetic and Askaryan emissions the footprint on ground is slightly asymmetric. 
This means that the amplitude of the radio signal is not a simple function of the distance to shower axis, unless it is corrected for the asymmetry \cite{KostuninTheory2015}. 
The size of the asymmetry corresponds to the strength of the Askaryan effect relative to the geomagnetic emission, which depends on the zenith angle and distance to the shower axis (figure \ref{fig_asymmetry}). 
Recently a phase delay of about $1\,$ns between both effects has been measured by LOFAR, which means that the resulting polarization is slightly elliptical \cite{ScholtenLOFAR_circPol2016}. 
Since the values in figure \ref{fig_asymmetry} had been determined before this discovery, it is not clear which values consider the full strength of the Askaryan emission and which values consider only the fraction of the Askaryan effect in phase with the geomagnetic emission.
Thus, some values might slightly underestimate the true fraction of Askaryan emission, which has to be determined in future studies.

Finally, also the slight asymmetry of the hyperbolic radio wavefront \cite{2014ApelLOPES_wavefront, CorstanjeLOFAR_wavefront2014} might be explained by the phase shift of both effects. 
The radio wavefront is of approximately hyperbolic shape, as expected for a finite line source, i.e., the radio wavefront approximates a cone at larger distances from the shower front. 
This cone has a typical angle to the shower plane (plane perpendicular to the shower axis) of $1-2^\circ$, where the value of the cone angle is approximately proportional to the distance of the shower maximum. 
Like for the asymmetry of the radio footprint, a slight asymmetry has also been predicted by simulations for the wavefront \cite{2014ApelLOPES_wavefront}.
This now seems understandable, since the pulse shape will depend on whether the Askaryan and geomagnetic emissions interfere destructively or constructively.
In CoREAS simulations made for LOPES it was seen that this cone is approximately $10\,\%$ steeper towards East than towards West, which seems to be in rough agreement with the size of the phase shift observed by LOFAR.

Consequently, as of today, all properties of the radio signal, i.e., its amplitude, polarization, and arrival time, seem to be understood at a level of around $10\,\%$, and are quantitatively reproducible by simulation codes.

\begin{table*}[p]
\centering
\caption{Selection of modern antenna arrays for air-shower or neutrino detection and references for further reading (more extended table in Ref.~\cite{SchroederReview2016}).} \label{tab_experiments}
\vspace{0.3cm}
\small
\begin{tabular}{lccccc}
\hline
Name of&Operation  & \multicolumn{2}{c}{aiming at} & Medium of &References\\
experiment& period & Cosmic Rays & Neutrinos & radio emission &\\
\hline
Yakutsk & since 1972 & x & & air & \cite{YakutskJETP2016}\\
LOPES & 2003 - 2013 & x & & air & \cite{FalckeNature2005, SchroederLOPESsummaryECRS2012}\\
CODALEMA& since 2003 &  x & & air & \cite{ArdouinBelletoileCharrier2005, Ardouin2009}\\
ANITA(-lite)& first flight 2004 & x & x & air + ice & \cite{ANITA_CR_PRL_2010, ANITA_CR_Energy_2016}\\
TREND & 2009 - 2014 & x & & air & \cite{TREND2011}\\
AERA & since 2010 & x & & air & \cite{RadioOffline2011, AERA_ICRC2015}\\
ARA& since 2010 &  & x & ice & \cite{ARA_2016}\\
LOFAR & since 2011 & x & x & air + moon & \cite{SchellartLOFAR2013, LOFARNature2016}\\
Tunka-Rex & since 2012 & x & & air & \cite{TunkaRex_NIM_2015, TunkaRex_Xmax2016}\\ 
ARIANNA& since 2012 & x & x & air + ice & \cite{ARIANNA_2015, ARIANNA_ARENA2016}\\
SKA-low & planned & x & x & air + moon & \cite{HuegeSKA_ICRC2015, JamesSKA_ICRC2015} \\
GRAND & planned & x & x & air + mountain & \cite{GRAND_ICRC2015} \\
\hline
\end{tabular}
\end{table*}

\begin{figure*}[p]
  \centering
  \includegraphics[width=0.99\linewidth]{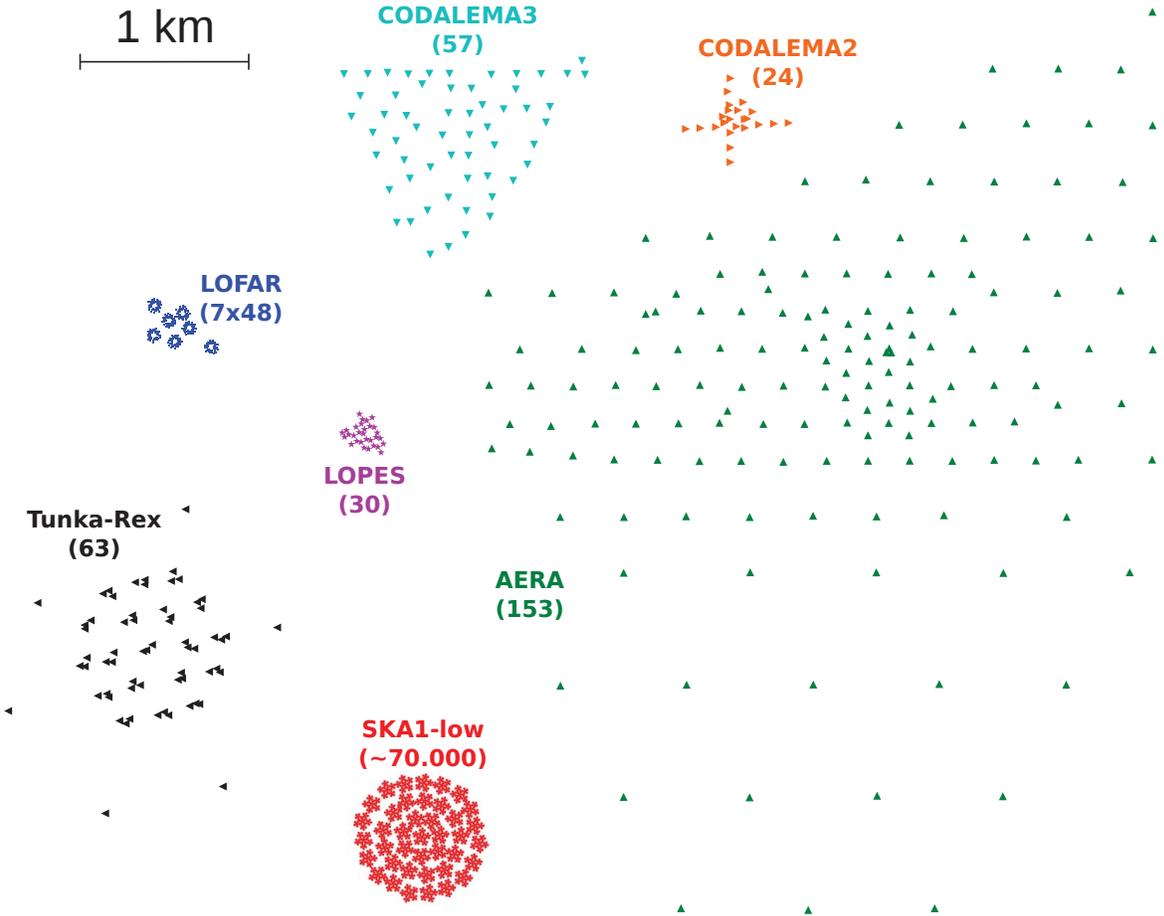}
  \caption{Layouts of selected antenna arrays for air showers. The number of antenna stations is given in brackets. For LOFAR the configuration of active antennas varies between observations, and typical air showers are measured by a few 100 antennas. SKA-low will be built in a few years, and it is not yet decided how many of the about $70,000$ antennas will be available for air-shower detection.}
  \label{fig_mapOfArrays}
\end{figure*}

\section{Radio arrays}
Radio signal from air-showers have first been measured in the 1960s by several analog antenna arrays \cite{jelley, Allan1971}. 
These historic measurements have been sufficient to discover the qualitative dependencies of the radio signal, but large quantitative uncertainties remained. 
Hence, the accuracy achieved for air-shower parameters was insufficient.
This changed in the 2000s when several digital experiments were built and sophisticated methods for computing-intensive data analysis could be used. 
In particular LOPES, a digital radio interferometer triggered by the KASCADE-Grande particle-detector array \cite{FalckeNature2005, Apel2010KASCADEGrande}, revived the field, and only recently its data analysis has been finished. 
At about the same time CODALEMA, an antenna array in a radio-quiet location in France, detected its first air showers \cite{ArdouinBelletoileCharrier2005}. 
A few years later, a second generation of experiments followed, e.g., the Auger Engineering Radio Array (AERA) \cite{SchroederAERA_PISA2015}, LOFAR \cite{SchellartLOFAR2013}, Tunka-Rex \cite{TunkaRex_NIM_2015}, and a few others (see table \ref{tab_experiments}). 

These second-generation arrays consist of the order of 100 antennas distributed over an area of a typical size of the order of square-kilometers (see figure~\ref{fig_mapOfArrays}), where LOFAR is leading in number of antennas, and AERA in size ($17\,$km$^2$).
These arrays use different antenna types, but similar frequency bands of about $30-80\,$MHz, which  provides a good signal-to-noise ratio for reasonable cost: 
at lower frequencies the Galactic noise is too strong, higher frequencies would require more expensive electronics and a denser antenna spacing.
In addition to advances in technology and in the theoretical understanding of the radio emission, the accurate calibration of these arrays has become crucial for the recent success of the radio technique: 
Current radio arrays became competitive with established techniques in accuracy only thanks to new methods for nanosecond-precise time synchronization \cite{SchroederTimeCalibration2010, AERAairplanePaper2015, CorstanjeLOFAR_timeCalibration2016}, and methods for amplitude calibration with absolute accuracies of better than $20\,\%$ for the field strength of the radio signal \cite{2015ApelLOPES_improvedCalibration, NellesLOFAR_calibration2015, TunkaRex_NIM_2015, AERAantennaPaper2012}.

Finally, there are additional detector concepts other than ground arrays, in particular the observation of inclined air-showers with single antenna stations on mountains, satellites, or balloons \cite{ANITA_CR_PRL_2010}, the observation of the lunar regolith \cite{BrayReview2016}, or the search for the radio signal of neutrino-induced showers in ice \cite{ARIANNA_2015, ARA_2016}.
For a more detailed overview, please see reference \cite{SchroederReview2016}.

\begin{figure*}[t]
  \centering
  \includegraphics[width=0.46\linewidth]{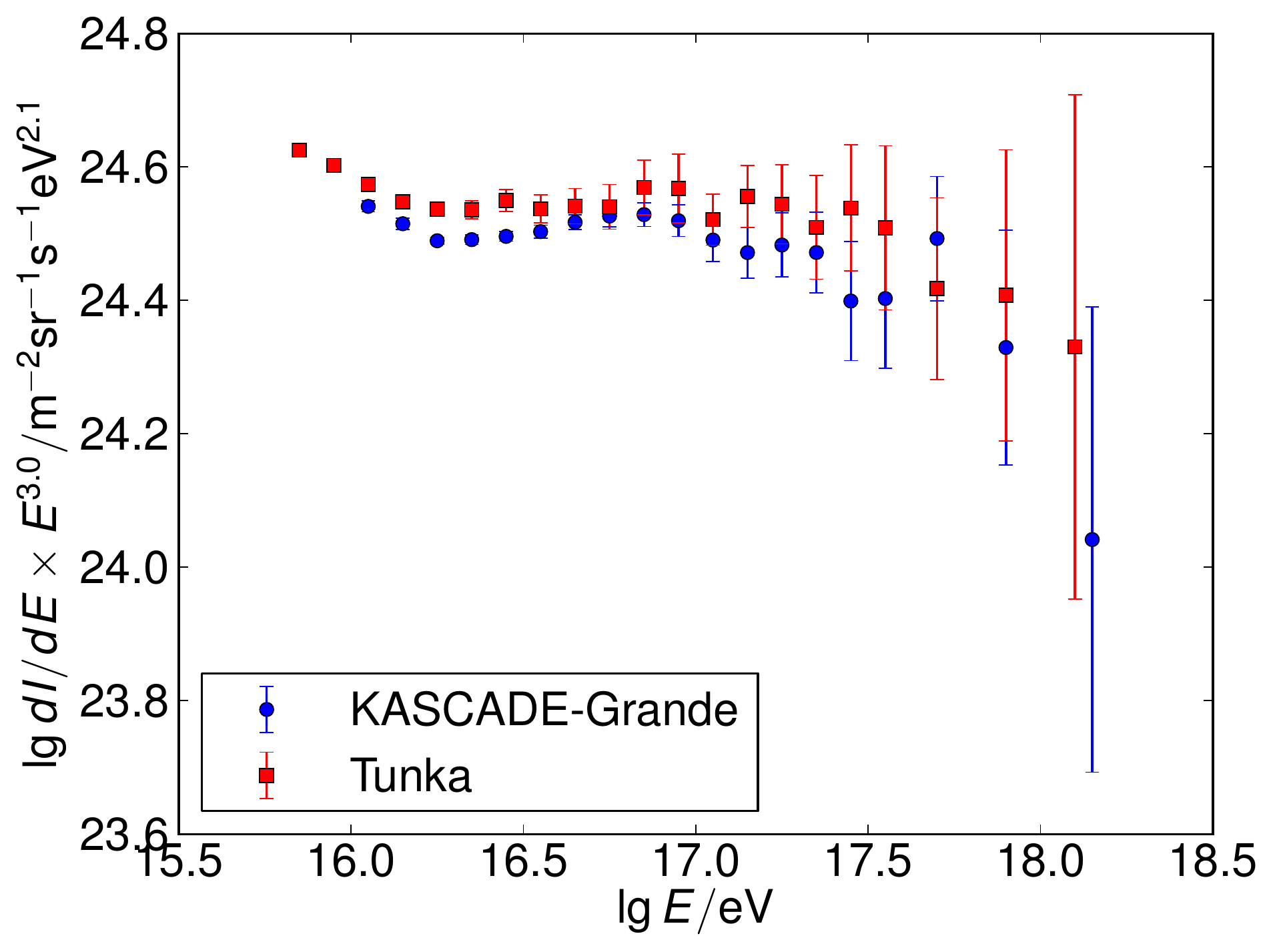}
  \hfill
  \includegraphics[width=0.51\linewidth]{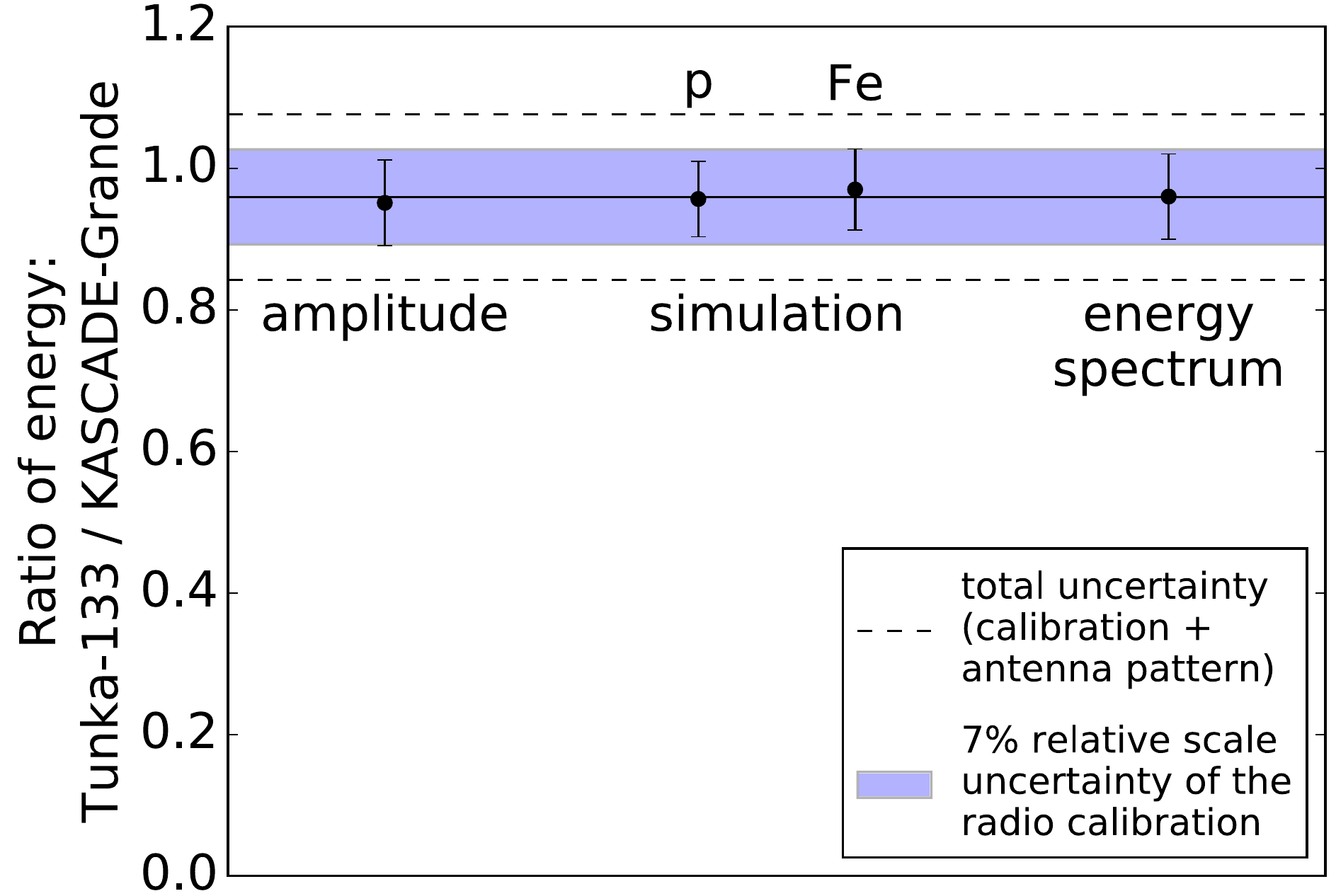}
  \caption{Comparison of the energy scales of KASCADE-Grande and Tunka-133 via their radio extensions LOPES and Tunka-Rex: a small scale shift between the energy spectra (left, \cite{2012ApelKGenergySpectrum,Tunka133_NIM2014}) is consistent with the scale difference measured with the radio antennas and evaluated by two different methods (right): first, a purely experimental method based on the measured radio amplitude per shower energy; second, a comparison of measurements and CoREAS simulations taking into account the situation and conditions of the experiments \cite{TunkaRexScale2016}.}
  \label{fig_energyScaleComparison}
\end{figure*}

\section{Current performance of the radio technique}
The precisions and scale accuracies reached by current radio arrays for the most important air-shower parameters are now similar to the ones achieved by the established methods. 
The accuracy for the arrival direction is better than $1^\circ$ \cite{2014ApelLOPES_wavefront, CorstanjeLOFAR_timeCalibration2016}, which is more than sufficient for charged cosmic rays. 
The experimentally demonstrated precision for the shower energy is about $15-20\,\%$, as has been shown in various comparisons of antenna arrays and their host detectors \cite{2014ApelLOPES_MassComposition, TunkaRex_Xmax2016, AERAenergyPRL2015}. 
Simulation studies predict that the energy precision in principle can be even better than $10\,\%$ \cite{2014ApelLOPES_MassComposition, KostuninTheory2015, GlaserTheoryPaper2016}. 
This might be achievable not only theoretically, but also in practice since the radio signal seems to depend less on atmospheric conditions than optical detection methods \cite{AugerGDAS_2012, AugerNIM2015}: 
first studies show that changes of the refractivity of the air, e.g., due to humidity, affect the radio signal on the level of a few percent, only \cite{ANITA_CR_Energy_2016, LOFARNature2016}.
This is slightly more than for particle measurements \cite{ParticleAtmosphericEffects_2009}, but still tolerable when aiming at $10\,\%$ total accuracy.
Currently, the scale accuracy is of the order of $15\,\%$ \cite{AERAenergyPRL2015}, and might be improved soon to below $10\,\%$ by better calibrations of the antennas. 
Lately, LOPES and Tunka-Rex have demonstrated that relative comparisons of the energy scale are already possible on a $10\,\%$ level, using exactly the same external calibration source for both experiments (see figure~\ref{fig_energyScaleComparison}, \cite{TunkaRexScale2016}).

Most important and most difficult is an accurate measurement of the mass composition as function of energy. 
Various parameters of the radio signal are sensitive to the atmospheric depth of the shower maximum, $X_\mathrm{max}$.
That is one of the best statistical estimators for the cosmic-ray composition, although its interpretation goes along with systematic uncertainties of hadronic interaction models \cite{KampertUnger2012}.
There is direct experimental evidence that the slope of the lateral distribution is sensitive to $X_\mathrm{max}$ \cite{2012ApelLOPES_MTD, TunkaRex_Xmax2016}, and there are convincing theoretical indications and experimental hints that also the following radio observables are sensitive to $X_\mathrm{max}$: 
the width of the radio footprint \cite{NellesLOFAR_measuredLDF2015}, the steepness of the hyperbolic radio wavefront \cite{2014ApelLOPES_wavefront}, and the slope of the frequency spectrum \cite{Grebe_ARENA2012}.

While the sparse and economic Tunka-Rex array has demonstrated a precision of $40\,$g/cm$^2$ for $X_\mathrm{max}$ with the lateral-slope method, the much denser LOFAR array reaches an accuracy of about $20\,$g/cm$^2$ by using a more sophisticated method implicitly taking into account all features of the radio footprint including both, its width and slope \cite{BuitinkLOFAR_Xmax2014, LOFARNature2016}.
This estimate for the accuracy already includes known systematic uncertainties, and is similar to the accuracy achieved by the leading fluorescence technique. 
Because of the small area of LOFAR, current fluorescence observations still have much larger statistics and exposure, despite of their limitation to dark and clear nights \cite{AugerHEATXmaxICRC2015}.
Nevertheless, these results show that it is no longer a principle question whether the radio technique can contribute to cosmic-ray physics, but just a question when sufficiently large and dense radio arrays will be built.

\section{Conclusion and Outlook}
Many air-shower arrays already feature an operating radio extension, which provides additional accuracy for relatively low cost. 
These radio arrays can achieve accuracies for the arrival direction, energy and position of the shower maximum comparable to those of the leading air-Cherenkov and air-fluorescence techniques. 
In future even higher accuracies might be possible, since the radio signal is well understood, depends little on atmospheric conditions, and calibrations and analysis techniques are constantly improving. 
Since additional accuracy is exactly what is needed to distinguish between different scenarios for the origin of the most energetic Galactic and extragalactic cosmic-rays, this provides a clear science case for radio extensions. 
In particular the combination of antennas with muon detectors seems promising, since the ratio between the number of muons and the size of the electromagnetic component measured by radio yields additional sensitivity for the mass composition, which is complementary to $X_\mathrm{max}$ \cite{Holt_TAUP2015}.

The threshold for the radio technique is around $10^{17}\,$eV depending only marginally on the antenna type, since the limiting factor is external Galactic radio noise. 
If full efficiency is desired for all arrival directions, then antenna arrays have to be comparably dense with spacings of the order of $100-200\,$m. 
Very dense arrays, like the SKA \cite{HuegeSKA_ICRC2015}, thus, will be able to measure air showers around $10^{17}\,$eV with unprecedented accuracy.
However, antenna spacings of $1-2\,$km seem to be reasonable at zenith angles around $75^\circ$, since the footprint becomes much larger for inclined showers \cite{AERA_ICRC2015}. 
In this sense the radio technique is very complementary to other techniques.
It provides a unique way to measure the electromagnetic component of such inclined showers. 
This will be exploited by the proposed GRAND experiment, which aims at the detection of tau neutrinos interacting in mountains, and simultaneously will have the world-leading exposure for ultra-high-energy cosmic rays \cite{GRAND_ICRC2015}.

While antennas seem to be perfect as extensions of particle detectors, radio as stand-alone technique is more challenging, in particular since radio disturbances are difficult to distinguish from real events without the complementary information of another detector. 
Still the proof-of-principle for radio as stand-alone technique has been made \cite{ANITA_CR_Energy_2016, ARIANNA_ARENA2016}.
Especially in radio-quiet environments, self-triggered radio extension seems to have a sufficient purity for cosmic particles.
Consequently, radio arrays will be a well-suited detector for high-energy neutrinos interacting in ice, and with ARA and ARIANNA two prototype arrays are operating at Antarctica \cite{ARIANNA_2015, ARA_2016}. 
Since radio waves are attenuated less than Cherenkov light, these radio arrays can be sparser and aim at higher energies than optical neutrino detection can do. 

In summary, the radio technique is the ideal complement to existing techniques for cosmic-ray and neutrino detection.
Future radio extensions and stand-alone experiments can bring the additional measurements required to finally understand how nature produces the most energetic particles in the universe.

\bigskip 
\begin{acknowledgments}
Thanks go to the conference organizers, to my colleagues at KIT, to the LOPES, Pierre Auger and Tunka-Rex Collaborations for fruitful discussions, to the Helmholtz Alliance for Astroparticle Physics (HAP), and to DFG for grant Schr 1480/1-1. 
\end{acknowledgments}

\bigskip 
\bibliography{ecrs2016.bib}

\end{document}